\documentstyle[prd,preprint,aps]{revtex}
\begin{document}
\draft
\title{Elastic neutrino-electron scattering and effective couplings \\
$(g^{\nu e}_V)_{LRSM}$ and $(g^{\nu e}_A)_{LRSM}$ in a Left-Right symmetric model}

\author{A. Guti\'errez-Rodr\'{\i}guez $^1$, M. A. Hern\'andez-Ru\'{\i}z $^1$ and M. Maya $^2$}

\address{(1) Facultad de F\'{\i}sica, Universidad Aut\'onoma de Zacatecas\\
Apartado Postal C-580, 98060 Zacatecas, Zacatecas M\'exico.}

\address{(2) Facultad de Ciencias F\'{\i}sico Matem\'aticas, Universidad Aut\'onoma de Puebla\\
Apartado Postal 1364, 72000, Puebla, Puebla M\'exico.}

\date{\today}
\maketitle
\begin{abstract}

We start from a Left-Right Symmetric Model with massive Dirac neutrinos,
with an electromagnetic structure that consists of a charge radius, and we
calculate the cross-section of the scattering $\nu_{\mu}e^-\rightarrow \nu_{\mu}e^-$.
Subsequently, we calculate the simultaneous contribution of the charge radius,
of the additional $Z_{2}$ heavy gauge boson and of the mixing angle $\phi$
parameter of the model at the constants of couplings $(g^{\nu e}_{V})_{LRSM}$
and $(g^{\nu e}_{A})_{LRSM}$. 
\end{abstract}


\narrowtext

\section{Introduction}
Neutrinos seem to be likely candidates for carrying features of physics beyond
the Standard Model (SM) \cite{S.L.Glashow}. Apart from masses and mixings,
charge radius, magnetic moments and electric dipole moments are also signs of new
physics, and are of relevance in terrestrial experiments, the solar neutrino
problem \cite{Voloshin,Vogel,Degrassi,Lucio}, astrophysics and cosmology \cite{Cisneros,Salati}.

At the present time, all the available experimental data for electroweak
processes can be well understood in the context of the Standard Model of the
electroweak interactions (SM) \cite{S.L.Glashow}, except the results of the
SUPER-KAMIOKANDE experiment on the neutrino mass \cite{Fukuda}. However,
Super-Kamiokande not is the only experiments in disagreement with the SM, also
are includen at GALLEX, SAGE, GNO, HOMESTAKE and LSND \cite{SAGE}. Hence, the
SM is the starting point for all the extended gauge models. In other words, any
gauge group with physical sense must have as a subgroup the
$SU(2)_{L} \times U(1)$ group of the standard model, in such a way that its
predictions agree with those of the SM at low energies. The purpose of the
extended theories is to explain some fundamental aspects which are not
clarified in the framework of the SM. One of these aspects is the origin of the
parity violation at the current energies. The Left-Right Symmetric Model (LRSM)
based on the $SU(2)_{R}\times SU(2)_{L}\times U(1)$ gauge group
\cite{J.C.Pati} gives an answer to that problem, since it restores the parity
symmetry at high energies and gives their violations at low energies as a
result of the breaking of gauge symmetry. Detailed discussions on LRSM can
be found in the literature \cite{J.C.Pati,R.N.Mohapatra,G.Senjanovic,G.Senjanovic1}.

Although in the framework of the SM, neutrinos are assumed to be electrically
neutral. Electromagnetic properties of the neutrino are discussed in many
gauge theories beyond the SM. Electromagnetic properties of the neutrino may
manifest themselves in a magnetic moment of the neutrino as well as in a
non vanishing charge radius, both making the neutrino subject to the electromagnetic
interaction.

The present paper is an extension of previous work by one of the authors
(M. Maya {\it et al.}), considering electromagnetic properties of the neutrino.
In this work, we start from a Left-Right Symmetric Model (LRSM) with
massive Dirac neutrinos left and right, with an electromagnetic structure
that consists of a charge radius and of an anomalous magnetic moment, and we
calculated the cross-section of the scattering $\nu_{\mu}e^{-}\rightarrow
\nu_{\mu}e^{-}$. The Feynman diagrams which contribute  to the process
$\nu_{\mu}e^{-}\rightarrow \nu_{\mu}e^{-}$ are shown in Fig. 1.

In previous papers \cite{Czakon}, possible corrections at the couplings
of the fermion with the gauge boson were calculated, in particular the couplings
$g_{V}$ and $g_{A}$ of leptons with the neutral boson $Z^{0}$, which have been
measured with great precision in LEP and CHARM II \cite{CHARM II}.
In this work we calculate the simultaneous contribution of the charge radius,
of the additional $Z_{R}$ gauge boson and of the angle $\phi$ parameter of
the LRSM at the constants of couplings $(g^{\nu e}_{V})_{LRSM}$ and $(g^{\nu e}_{A})_{LRSM}$.
The charge radius in the LRSM considered is simply treated as a new parameter.
One is thus dealing with a purely phenomenological analysis.

For an analysis of the electromagnetic form factors of the neutrino from a point
of see theoretical in left-right models, see for example the Ref. \cite{Branco}.

This paper is organized as follows. In Sect. II we carry out the calculus of the
process $\nu_{\mu}e^{-}\rightarrow \nu_{\mu}e^{-}$. In Sect. III we present the
expressions for the constants of couplings $(g^{\nu e}_{V})_{LRSM}$ and
$(g^{\nu e}_{A})_{LRSM}$. In Sect. IV we achieve the numerical computations.
Finally, we summarize our results in Sect. V.

\section{The neutrino-electron scattering}

We will assume that a massive Dirac neutrino is characterized by two phenomenological
parameters, a magnetic moment $\mu_{\nu}$, expressed in units of the electron
Bohr magnetons, and a charge radius $\langle r^{2}\rangle$. Therefore, the expression
for the amplitude $\cal M$ of the process $\nu_\mu e^-\rightarrow \nu_\mu e^-$
due only to $\gamma$ and $Z^0$ exchange, according to the diagrams depicted in
Fig. 1 is given by

\begin{equation}
{\cal M}_\gamma =\bar \nu (k_2)\frac{\Gamma ^\mu}{q^2}\nu (k_1)\bar e(p_2)\gamma _\mu e(p_1),
\end{equation}

\noindent with

\begin{equation}
\Gamma ^\mu = eF_1(q^2)\gamma ^\mu -\frac{e}{2m_\nu}F_2(q^2)\sigma ^{\mu \nu}q_\nu,
\end{equation}

\noindent the neutrino electromagnetic vertex, where $q$ is the momentum
transfer and $F_{1,2}(q^2)$ are the electromagnetic form factors of the neutrino.
Explicitly \cite{Vogel}

\[
F_1(q^2)=\frac{1}{6}q^2 \langle r^2 \rangle, 
\]
\[
F_2(q^2)=-\mu _\nu \frac{m_\nu}{m_e}, 
\]

\noindent where as already are mentioned $\langle r^2 \rangle$ is the neutrino
mean-square charge radius and $\mu_\nu$ the anomalous magnetic moment.

\noindent Furthermore

\begin{eqnarray}
{\cal M}_{Z^0}&=&\frac{G_F}{\sqrt{2}}[P\bar \nu(k_2)\gamma^\mu \nu(k_1)\bar e(p_2)\gamma_\mu e(p_1)
             +Q\bar \nu(k_2)\gamma^\mu \gamma_5 \nu(k_1)\bar e(p_2)\gamma_\mu e(p_1)\nonumber\\
           &&+R\bar \nu(k_2)\gamma^\mu \nu(k_1)\bar e(p_2)\gamma_\mu \gamma_5 e(p_1)
             +S\bar \nu(k_2)\gamma^\mu \gamma_5 \nu(k_1)\bar e(p_2)\gamma_\mu \gamma_5 e(p_1)],
\end{eqnarray}

\noindent where

\begin{eqnarray}
P&=&(a+2b+c)g_V,\nonumber\\
Q&=&(-a+c)g_A,\nonumber\\
R&=&(-a+c)g_V,\\
S&=&(a-2b+c)g_A\nonumber.
\end{eqnarray}

The constants $a$, $b$ and $c$ \cite{Ceron} depend only of the parameters of
the LRSM, and are

\begin{eqnarray}
a&=&(c_{\phi}-\frac{s^{2}_{W}}{r_{W}}s_{\phi})^{2}+\gamma(\frac{{s^{2}_{W}}}{r_{W}}c_{\phi}+s_{\phi})^{2},\nonumber\\
b&=&(c_{\phi}-\frac{{s^{2}_{W}}}{r_{W}}s_{\phi})(-\frac{{c^{2}_{W}}}{r_{W}}s_{\phi})
   +\gamma(\frac{{s^{2}_{W}}}{r_{W}}c_{\phi}+s_{\phi})(\frac{{c^{2}_{W}}}{r_{W}}c_{\phi}),\\
c&=&(\frac{c^{2}_{W}}{r_{W}}s_{\phi})^{2}+\gamma(\frac{c^{2}_{W}}{r_{W}}c_{\phi})^{2}, \nonumber\\
\gamma&=&(\frac{M_{Z_{1}}}{M_{Z_{2}}})^{2}=(\frac{M_{Z_L}}{M_{Z_R}})^2,\nonumber
\end{eqnarray}

\noindent where $M_{Z_{1}}$ and $M_{Z_{2}}$ are the masses of the neutral bosons
that participate in the interaction. $\gamma$ together with $\phi$ are the two
new parameters that are introduced in the LRSM. While $g_V=-\frac{1}{2}+2\sin^2\theta_W$
and $g_A=-\frac{1}{2}$, according to the experimental data \cite{Review}.

The square of the amplitude is obtained by sum over spin states of the final fermions, so

\begin{equation}
\sum_{sp}|{\cal M}_{T}|^2=\sum_{sp}(|{\cal M}_\gamma|^2
                          +|{\cal M}_{Z^0}|^2+{\cal M}_{Z^0 }{\cal M}^{\dagger}_\gamma
                          +{\cal M}^{\dagger}_{Z^0} {\cal M}_\gamma),
\end{equation}

\noindent where:

\begin{eqnarray}
\sum_{sp}|{\cal M}_\gamma|^2&=&\frac{8}{9}e^4{\langle r^2 \rangle}^2 E^4 (5+2x+x^2),\\
\sum_{sp}|{\cal M}_{Z^0}|^2&=&16G^2_{F}E^4[(P^2+Q^2+R^2+S^2)(5+2x+x^2)\nonumber\\
                           && +2(PS+QR)(3-2x-x^2)],\\
\sum_{sp}({\cal M}_{Z^0}{\cal M}^{\dagger}_{\gamma} + {\cal M}^{\dagger}_{Z^0} {\cal M}_\gamma )&=& \frac{32}{3\sqrt{2}}
                           e\langle r^2 \rangle G_F E^4 [5P+3S+(P-S)(2x+x^2)],
\end{eqnarray}

\noindent and $x=\cos\theta$, with $\theta$ the scattering angle.

In the expressions (7), (8) and (9) it is observed that there is no contribution
of the anomalous magnetic moment; this is due to the fact that the magnetic moment induces
change of helicity, which is not considered here. However, there is contribution
of the electroweak charge radius, of the heavy gauge boson $Z_R$ and of the
mixing angle $\phi$.

The scattering cross-section in the center of mass system (where $s$ is the
square of the center-of-mass energy) is given by

\begin{equation}
\sigma=\int \frac{d\Omega}{64\pi^2 s}\frac{1}{2}\sum_{sp}|{\cal M}_T|^2,
\end{equation}

\noindent where the square of the total amplitude of transition ${\sum}_{sp}|{\cal M}_T|^2$
is given in the Eq. (6).

We write the total cross-section and the interference cross-section of the
reaction $\nu_\mu e^- \rightarrow \nu_\mu e^-$ in the laboratory system
using the relation \cite{Review}

\begin{equation}
\frac{E^4}{s}=\frac{m_eE_{lab}}{8},
\end{equation}

\noindent we obtain the following

\begin{eqnarray}
\sigma^{LRSM}_T &=& \frac{G^2_Fm_eE_\nu}{2\pi}\{2{\cal R}^{2}+2{\cal R}(P+S)+\frac{(P+S)^2+(Q+R)^2}{2}\nonumber\\
                && +\frac{1}{3}[2{\cal R}^{2}+2{\cal R}(P-S)+\frac{(P-S)^2+(Q-R)^2}{2}]\},
\end{eqnarray}

\noindent while for the interference:

\begin{equation}
\sigma^{LRSM}_{\gamma Z^0}=\frac{2G^2_Fm_eE_\nu}{3\pi}{\cal R}(2P+S),
\end{equation}

\noindent where 

\begin{equation}
{\cal R}=\frac{\sqrt{2}\pi \alpha}{3G_F}\langle r^2 \rangle,
\end{equation}

\noindent is defined, and $\alpha=\frac{e^2}{4\pi}$ is the fine-structure
constant. P, Q, R and S are defined in Eq. (4) and to be in terms of the
LRSM parameters $(\phi, M_{Z_R})$. Let us rewrite the interference cross-section
of the following form:

\begin{eqnarray}
\sigma^{LRSM}_{\gamma Z^0}&=&K{\langle r^2 \rangle}(2g_V+g_A)
[\{1+(\frac{2g_V a_W+g_A b_W}{2g_V+g_A})\gamma \}c^2_\phi \nonumber\\
 && +\{1+(\frac{2g_V a_W+g_A b_W}{2g_V+g_A})\gamma \}s^2_\phi
+(\gamma -1)(\frac{2g_V+g_A b^{'}_{W}}{2g_V+g_A})s_\phi c_\phi],
\end{eqnarray}

\noindent where
\[
K=\frac{4G_Fm_e\alpha E_\nu}{9\sqrt{2}},
\]

\noindent and

\[
a_W=\frac{(s^2_W+c^2_W)^2}{r^2_W}, \hspace*{5mm}  b_W=\frac{(s^2_W-c^2_W)^2}{r^2_W},
\hspace*{5mm} b^{'}_W=\frac{2(s^2_W-c^2_W)}{r_W},
\]

\noindent the contribution of the new heavy boson $Z_R$ in each term of the Eq. (15)
stays patent. In the limit $M_{Z_R}\rightarrow \infty$, $\gamma \rightarrow 0$
and $\phi = 0$ one has the charge radius that is denoted by ${\langle r^2 \rangle}_0$,
and the interference cross-section is

\begin{equation}
\sigma^0_{\gamma Z^0}=K{\langle r^2 \rangle}_0 (2g_V+g_A),
\end{equation}

\noindent which agrees with the term of interference of the Ref. \cite{Vogel,Salati}.
Here $\langle r^2 \rangle _0$ is the charge radius inside of the SM minimally
extended to include massive Dirac neutrinos, with an electromagnetic structure.

\section{The vector-axial couplings}

At the present time, the most precise direct determinations of $g^{\nu e}_{V,A}$
come from the CHARM II experiment using $\nu_\mu e^-$ scattering \cite {CHARM II}
$g^{\nu e}_{V}=-0.035\pm 0.017$ and $g^{\nu e}_{A}=-0.503\pm 0.017$ at $1\sigma$,
in agreement with the SM. In this section, we obtain expressions for $(g^{\nu e}_{V})_{LRSM}$
and $(g^{\nu e}_{A})_{LRSM}$ in terms of the SM couplings $g^{SM}_{V}$ and $g^{SM}_{A}$,
of the electroweak charge radius, the mixing angle $\phi$ parameter of the
LRSM and of the heavy boson $Z_{R}$. Returning to the expression of the total
cross-section Eq. (12) and defined

\[
P=f_1g^{\nu e}_V, \hspace*{5mm} Q=-f_2g^{\nu e}_V, \hspace*{5mm} R=-f_2g^{\nu e}_A, \hspace*{5mm} S=f_3g^{\nu e}_A, 
\]

\noindent we obtain

\begin{eqnarray}
\sigma^{LRSM}_T&=&\frac{G^{2}_Fm_eE_\nu}{2\pi}\{{\cal R}^2+2{\cal R}(f_1g^{\nu e}_V+f_3g^{\nu e}_A)
               +\frac{1}{2}(f_1g^{\nu e}_V +f_3g^{\nu e}_A)^2
               +\frac{1}{2}f^{2}_2 (g^{\nu e}_V+g^{\nu e}_A)^2\nonumber\\
               && +\frac{1}{3}[{\cal R}^2+2{\cal R}(f_1g^{\nu e}_V-f_3g^{\nu e}_A)
               +\frac{1}{2}(f_1g^{\nu e}_V-f_3g^{\nu e}_A)^2
               +\frac{1}{2}f^{2}_2(g^{\nu e}_V-g^{\nu e}_A)^2]\},
\end{eqnarray}

\noindent where

\begin{equation}
f_1=u^2+\frac{v^2}{r^2_W}\gamma, \hspace*{5mm} f_2=uv(1-\gamma), \hspace{5mm} f_3=v^2+u^2r^2_W\gamma,
\end{equation} 

\noindent and

\[
u=\cos\phi-\frac{\sin \phi}{r_W}, \hspace*{5mm} v=\cos\phi+r_W\sin\phi
\]

\noindent with ${\cal R}$ and $\gamma$ as defined in the Eqs. (5) and (14).

The total cross-section for the reaction $\nu_\mu e^-\rightarrow \nu_\mu e^-$
in the context of the SM \cite{Review} is

\[
\sigma^{SM}=\frac{G^2_Fm_eE_\nu}{2\pi}\{(g^{\nu e}_V+g^{\nu e}_A)^2
            +\frac{1}{3}(g^{\nu e}_V-g^{\nu e}_A)^2\},
\]

\noindent likewise we have

\begin{equation}
\sigma^{LRSM}_T=\frac{G^{2}_Fm_eE_\nu}{2\pi}\{((g^{\nu e}_V)_{LRSM}+(g^{\nu e}_A)_{LRSM})^2
              +\frac{1}{3}((g^{\nu e}_V)_{LRSM}-(g^{\nu e}_A)_{LRSM})^2\}.
\end{equation}

From Eqs. (17) and (19), the couplings $(g^{\nu e}_V)_{LRSM}$ and $(g^{\nu e}_A)_{LRSM}$
in terms of the couplings of the SM and of the LRSM parameters are

\begin{equation}
(g^{\nu e}_V)_{LRSM}={\cal R}f_3+\frac{1}{2}(f_1f_3+f^{2}_2)g^{SM}_V,
\end{equation}

\begin{equation}
(g^{\nu e}_A)_{LRSM}=\{{\cal R}f_3+\frac{1}{2}(f_1f_3+f^{2}_2)\}g^{SM}_A.
\end{equation}

In these expressions, in the limit when $M_{Z_R}\rightarrow \infty$,
$\gamma \rightarrow 0$ and $\phi =0$, the couplings of the SM are recovered.

\section{Results}

In this section numerical results obtained by using the SM of the electroweak
interaction \cite{S.L.Glashow} are presented. We take $M_{Z_L}=91.187 \pm 0.007$ $GeV$,
$\sin^2\theta_W =0.2312$ and $g_V =-\frac{1}{2}+2\sin^2\theta_W$,
$g_A=-\frac{1}{2}$ according to the experimental data \cite{Review}.

The reported bounds for $M_{Z_R}$ \cite{Czakon,CHARM II,Review,J.Polak1} indicate
that this is not less than 100 $GeV$. We have chosen $M_{Z_R}\leq 800$ $GeV$ to
estimate the contribution of the additional heavy boson $M_{Z_R}$.

In Fig. 2 we have plotted $(g^{\nu e}_V)_{LRSM}$ from Eq. (20), as a function of
the LRSM parameters. We take $\phi=0$, ${{\cal R}< 0.02}$ \cite{CHARM II} and
we choose $M_{Z_R}\leq 800$ $GeV$. In this figure we observed that the experimental
value $g^{exp}_V=-0.052$ is reached for small values of ${\cal R}$ and large values
of $M_{Z_R}$. We obtain $-0.02\leq(g^{\nu e}_V)_{LRSM}\leq -0.05$ at the 90 $\%$ C. L..

In Fig. 3 we have plotted $(g^{\nu e}_A)_{LRSM}$ from Eq. (21) as a function of
${\cal R}$ and $M_{Z_R}$. The range of variation for the LRSM parameters is the
same as in Fig. 2. In this case the experimental data, $g^{exp}_A=-0.52$,
is reached for large values of ${\cal R}$ and $M_{Z_R}$. We obtain
$-0.5\leq(g^{\nu e}_A)_{LRSM}\leq -0.515$ at the 90 $\%$ C. L..

Fig. 4 shows the effective coupling $(g^{\nu e}_V)_{LRSM}$ as a function
of the LRSM parameters $\phi$ and $M_{Z_R}$. According  to the experimental
data, the allowed range for the mixing angle between $Z^{0}_L$ and $Z^{0}_R$
is $-0.009 \leq \phi \leq 0.004$ with a 90 $\%$ C.L. \cite{J.Polak1,L3,Maya},
we choose $M_{Z_R}\leq 800$ where $\phi$ is measured in radians and $M_{Z_R}$
in $GeV$. In the figure it is observed that the experimental data $g^{exp}_V=-0.052$
is reached for small values of $\phi$ and $M_{Z_R}$, while the SM data $g^{SM}_V=-0.038$
is reached for large values of $\phi$ and $M_{Z_R}$. We obtain
$-0.039\leq(g^{\nu e}_V)_{LRSM}\leq -0.044$ at the 90 $\%$ C. L.. Fig. 5 shows
the coupling $(g^{\nu e}_A)_{LRSM}$ as a function of $\phi$ and $M_{Z_R}$. The
range of variation for the LRSM parameters is the same as in Fig. 4. In this case
the experimental data $g^{exp}_A=-0.52$ is reached for small values of $\phi$
and $M_{Z_R}$, and the data SM $g^{SM}_V=-0.5$ is reached for large values of
both parameters. We obtain $-0.497\leq(g^{\nu e}_A)_{LRSM}\leq -0.51$ at the 90 $\%$ C. L..

Figs. 6 and 7 show the contribution of the LRSM parameters at the non-standard
couplings $(g^{\nu e}_V)_{LRSM}$ and $(g^{\nu e}_A)_{LRSM}$ to $M_{Z_R}=410$
$GeV$ \cite{Czakon,CHARM II,Review,J.Polak1} and $0\leq {\cal R} \leq 0.02$,
$-0.009\leq \phi \leq 0.004$. In the first case, the experimental result for $g_V$
is obtained for large ${\cal R}$ and $\phi$, while for $g_A$ the data experimental
is obtained for small ${\cal R}$ and $\phi$. In this case we obtain
$-0.028\leq(g^{\nu e}_V)_{LRSM}\leq -0.05$ and $-0.493\leq(g^{\nu e}_A)_{LRSM}\leq -0.51$
at the 90 $\%$ C. L..

\section{Conclusions}

In summary, we have determined an expression for (a) total cross-section;
(b) non-standard $\nu_\mu e^-$ vector and axial couplings. In case (a),
we find that for the cross-section of interference Eq. (15), the contribution of
the new boson $M_{Z_R}$ is evident. The SM prediction is obtained when we take
the limit $M_{Z_R}\rightarrow \infty$, $\gamma \rightarrow 0$ and $\phi = 0$ obtaining the
Eq. (16), which agrees with the term of interference report in the literature
Ref. \cite{Vogel,Salati}.

In case (b), the non-standard couplings $(g^{\nu e}_A)_{LRSM}$ and $(g^{\nu e}_V)_{LRSM}$
Figs. 2-7 are affected very sensitively by the LRSM parameters as well as for the charge
radius $\langle r^2\rangle$ which is implicit in the parameter ${\cal R}$. Our conclusion
is that the experimental value of the coupling $g_V$ favours small values of ${\cal R}$,
while the coupling $g_A$ points at large values of ${\cal R}$.

In this paper we point out the importance of continuing to examine the $\nu_\mu e^-$
scattering using a Left-Right Symmetric Model and assuming that a massive Dirac
neutrino is characterized by two phenomenological parameters, a magnetic moment
$\mu_\nu$, and a charge radius $\langle r^2\rangle$. It is also necessary to bear
in mind that the sensitivity of the experiment is increased, which must occur
in a next generation of accelerators or of a perfection of the current detectors
(SUPER-KAMIOKANDE) \cite{Fukuda}, and arguments are also given which make us think
that experiments with reactors as radioactive sources of neutrinos such as the
BOREXINO detector \cite{Ianni} and the reactor MUNU \cite{Broggini} can give answers
to the question of the charge radius, non-standard vector-axial couplings and
other parameters.

\newpage

\begin{center}
{\bf Acknowledgments}
\end{center}

This work was supported in part by {\it Consejo Nacional de Ciencia y Tecnolog\'{\i}a}
(CONACyT), {\it Sistema Nacional de Investigadores} (SNI) (M\'exico) and Programa
de Mejoramiento al Profesorado (PROMEP). A.G.R. would like to thank the organizers
of the  Summer School in Particle Physics 2001, Trieste Italy for their hospitality.

\newpage

\begin{center}
FIGURE CAPTIONS
\end{center}

\bigskip

\noindent {\bf Fig. 1} The Feynman diagrams contributing to the process
$\nu_{\mu}e^{-}\rightarrow \nu_{\mu}e^{-} $, in a left-right symmetric model.

\bigskip

\noindent {\bf Fig. 2} Plot of $(g^{\nu e}_V)_{LRSM}$ as a function of the LRSM
parameters $0\leq {\cal R}\leq 0.02$, $M_{Z_R}\leq 800$ $GeV$ with
$\phi =0$.

\bigskip

\noindent {\bf Fig. 3} Plot of $(g^{\nu e}_A)_{LRSM}$ as a function of the LRSM
parameters $0\leq {\cal R}\leq 0.02$, $M_{Z_R}\leq 800$ $GeV$ with
$\phi =0$.

\bigskip

\noindent {\bf Fig. 4} Same as in Fig. 2, but with $-0.009 \leq \phi \leq 0.004$,
$M_{Z_R}\leq 800$ $GeV$ and ${\cal R}=0.018$.

\bigskip

\noindent {\bf Fig. 5} Same as in Fig. 3, but with $-0.009 \leq \phi \leq 0.004$,
$M_{Z_R}\leq 800$ $GeV$ and ${\cal R}=0.018$.

\bigskip

\noindent {\bf Fig. 6} Same as in Fig. 2, but with $-0.009 \leq \phi \leq 0.004$,
$0\leq {\cal R}\leq 0.02$ and $M_{Z_R}=410$ $GeV$.

\bigskip

\noindent {\bf Fig. 7} Same as in Fig. 3, but with $-0.009 \leq \phi \leq 0.004$,
$0\leq {\cal R}\leq 0.02$ and $M_{Z_R}=410$ $GeV$.

\end{document}